\documentclass{aa}
\usepackage{graphics}
\begin{document}
\title{A method of estimation of the dynamical age of FR~II-type radio sources
from multi-frequency data}
\author{J. Machalski$^{1}$, K.T. Chy\.{z}y$^{1}$, \L . Stawarz$^{1,2}$ and D. Kozie\l $^{1}$}
\offprints{J. Machalski}
\institute{$^{1}$Astronomical Observatory, Jagellonian University,
ul. Orla 171, PL-30244 Cracow, Poland\\
$^{2}$ Landessternwarte Heidelberg, K\"onigstuhl, and MPI f\"ur Kernphysik, Saupfercheckweg 1, 69117 Heidelberg, Germany\\
email: machalsk@oa.uj.edu.pl}
\date{Received ....; accepted ...}
\abstract{Determination of ages of powerful radio sources is crucial in understanding galaxy evolution,
the activity cycle of galactic nuclei and their influence on the surrounding intergalactic medium.
So far, several different methods for age estimates of classical double radio galaxies have been proposed
and widely used in the literature, although each of them faces some difficulties due to observational
limitations and/or freedom in choosing the underlying model assumptions.}{We propose a new approach in
determining ages of FR~II type radio sources, on one hand exploiting a dynamical model developed for
these objects by Kaiser et al. (1997), and on the other hand using multifrequency radio observations not
necessarily restricted to the high-resolution ones.}{In particular, we apply the assumed dynamical model
to a number of FR~II type radio galaxies observed at different radio frequencies, and fit --- for each
 frequency separately --- the model free parameters to the observed sources' quantities. Such a procedure,
using enlarged in fact a number of observables, enables us to determine relatively precise ages and other
crucial characteristics (like the jets' kinetic power) of the analyzed sources.}{The resulting age estimates
agree very well with those obtained by means of `classical' spectral ageing method for objects 
not older than 10 Myr, for which good-quality spectral data are available. The presented method is however
also applicable in the case of the sources older than this, and/or the ones for which the only available
low-resolution radio data do not allow for detailed spectral ageing studies. Interestingly, the estimated 
ages always correspond to the realistic values of the jets' advance velocity $\sim 0.01\,c - 0.1\,c$}{Our
analysis indicates that the main factor precluding precise age determination for FR~II type radio
galaxies regards the poorly known shape of the initial electron energy distribution injected by the jet
terminal shocks to the expanding lobes/cocoons. We briefly consider this issue, and conclude that the
broad-band single power-law form assumed here may be accurate enough for the presented estimates, although
most likely it does not strictly correspond to some well-defined realistic particle acceleration process.
Instead, it should be considered as a simplest model approximation of the initial electron continuum, averaged
over a very broad energy range and over the age of the source, with the \emph{effective} spectral
index which may be different for different sources, however within the relatively narrow range
$p = 2.0-2.4$ suggested by our modeling.}
\keywords{galaxies: active -- galaxies: evolution -- radio continuum: galaxies}
\authorrunning{J. Machalski et al.}
\titlerunning{Dynamical age of FR~II-type radio sources}
\maketitle

\section{Introduction}

The dynamical age of extragalactic radio sources is of a great importance for studying
and understanding details of the physical processes taking place in both Active Galactic
Nuclei (AGNs) emitting energy in forms of jets boring through the ambient intergalactic
medium (IGM), and also lobes or cocoon consisting of shocked (backflowing) jet material
and shocked IGM. The `standard model' of classical double radio sources (Blandford \& 
Rees 1974; Scheuer 1974) implies that the axial lengthening of the cocoon is governed by
the balance between the jet's thrust and the ram pressure of the IGM, whilst the width of
the cocoon is determined by its internal pressure. This scenario has formed the basis for
all further analytical models of the dynamics and radio-emission properties of powerful
double radio sources (e.g., Begelman \& Cioffi 1989; Falle 1991; Nath 1995; Kaiser et al.
1997; Chy\.{z}y 1997; Blundell et al. 1999; Manolakou \& Kirk 2002, Kino \& Kawakatu 2005).

All such dynamical models allow one to predict a number of observational parameters of the
sources (e.g., linear size, width of the cocoon, radio power at different wavelengths) for
their given ages and redshifts, assuming values of the models' basic free parameters
(e.g., jet power, external medium density and its gradient, etc.). A comparison of
the predictions of the three main scenarios, i.e. the ones by Kaiser et al. (1997),
Blundell et al. (1999) and Manolakou \& Kirk (2002), was recently undertaken by Barai \&
Wiita (2005). They argued that {\sl no} existing model could give acceptable fits to the
common properties of radio sources from the survey considered (3CRR, 6CE, 7CRS), though the
Manolakou \& Kirk model gave `better overall results than do either of the other two'.   

The age of a classical double radio source can, in principle, be determined from its
apparent radio spectrum. There is no doubt that a radio continuum spectrum in different
parts of a radio source should contain crucial information about its history, depending
on various energy losses (and gains) of the radiating particles. This simple idea, relying
on the determination of a single characteristic break frequency in the radio spectrum
(at a given part of the source) caused by the above time-dependent losses, was used to
estimate the radiative ages and expansion speeds in large samples of powerful 3CR sources
(Myers \& Spangler 1985; Alexander \& Leahy 1987; Leahy et al. 1989; Carilli et al. 1991;
Liu et al. 1992), as well as in samples of low-luminosity and intermediate-luminosity radio
galaxies (Klein et al. 1995; Parma et al. 1999). However, the spectral break and steepening of
the spectrum beyond this break need not be entirely due to radiative ageing. A possible role
of the local magnetic field structure and evolution, details of the backflow process of
the cocoon material, or the difficulties in disentangling various energy losses of the
radiating particles have been raised in a number of papers (e.g. Rudnick et al. 1994; Eilek
\& Arendt 1996; Jones et al. 1999).

The task of resolving some of the above difficulties was undertaken by Kaiser (2000; hereafter
referred to as K2000) who has constructed a 3-dimensional model of the synchrotron
emissivity for a cocoon of a FR~II-type radio source, and applied it to the self-similar
model of the cocoon's dynamics developed by Kaiser \& Alexander (1997; hereafter referred
to as KA) and its extension by Kaiser et at. (1997; hereafter KDA). In the K2000 model,
the cocoon is split into small volume elements whose surface-brightness evolution is traced
individually. The author has argued that the proposed careful analysis of the bulk backflow
process and of the electron energy losses (both radiative and adiabatic) can provide an accurate
estimate of the sources' spectral ages, being in addition in agreement with the estimates of
their dynamical ages. Blundell \& Rawlings (2000) have contended however that this could only
be the case if these ages were much less than 10 Myr.

Analyzing the K2000 model, we realized that in the majority of the extended FR~II-type
radio sources --- even those without distorted lobe structures --- the surface-brightness
profiles are far from the expected smooth shapes, causing the fitted free parameters
to be highly uncertain. In addition, the K2000 method requires rather high resolution
observations of the radio lobes. But usually high-resolution observations of low-brightness
sources cause a serious loss of the flux density, so that application of the K2000 method is
in fact limited to a few strongest sources such as Cyg\,A.

The KDA model has been successively used by Machalski et al. (2004a, b) to derive the basic
physical parameters of the FR~II-type radio sources with known age, namely the jet power,
the central density of the ambient medium, the cocoon's energy density and pressure, and the total
energy deposited in the radio source. All of these values were derived by means of fitting
the model free parameters to the observed parameters of the source: its redshift, monochromatic
radio luminosity at 1.4 GHz, projected linear size, and the cocoon's axial ratio. In this paper
we propose a simplified method of the age determination for powerful radio sources based on a
further exploitation of the KDA model, which avoids the limitations present in the K2000 method.
In particular, we show that performing the above fitting procedure for a {\sl few} observing
frequencies, i.e. enlarging the number of the observables used by adding the radio spectrum data
of the entire cocoon, a {\sl unique} set of the age and the values for the two fundamental
model parameters --- the jet power and the central density of the ambient medium --- can be
found. This, in turn, enables us to determine relatively precisely the age of the analyzed source,
even if the available (low-resolution) radio data do not allow for detailed spectral ageing
studies. 

The paper is organized as follows. First, a brief presentation of the analytical KDA model
together with its application to the dynamical ageing studies are given in Section 2. In
Section 3 we present the dataset used in the subsequent analysis. Exemplary results of the
dynamical age determination for the radio galaxy Cygnus~A and its dependence on the assumed model
parameters are given in Section 4. Application of the proposed method to a variety of other FR~II
type radio sources is summarized in Section 5. Remarks on the implications of the obtained results for
modeling the electron injection spectrum in lobes of powerful radio sources are given in Section 6.
Final conclusions are presented in Section 7.

\section{The KDA model and its application}

\subsection{Source dynamics}

The dynamical expansion of a FR~II-type radio source built in the KDA model is based 
on the dynamics described in the self-similar KA model. Based on the original idea
of Blandford \& Rees (1974) and Scheuer (1974), it assumes that the classical double 
radio structure is formed by twin jets emerging from the AGN in opposite directions 
into a surrounding tenuous environment. The jets terminate in strong shocks 
where the jet particles are shocked, accelerated, and finally inflate the cocoon. 
A density distribution of the unperturbed external gas is approximated by a power-law 
radial distribution scaling with distance $r>a_{0}$ from the center of the host galaxy as

\[\rho(r)=\rho_{0}\left(\frac{r}{a_{0}}\right)^{-\beta},\]

\noindent
where $\rho_{0}$ is the central density at the core radius $a_{0}$, and the exponent 
$\beta$ describes the density profile in the simplified King's (1972) model. Such a 
distribution of the ambient medium is assumed to be invariant with redshift.

Half of the cocoon (from the core to the hotspot) is approximated by a cylinder of
length $L_{\rm jet}=D_{\rm s}/2$ and base diameter $b$. Here $D_{\rm s}$ is the
total deprojected length (linear size) of the source. The ratio of the source's length
to the base diameter is called the `axial ratio' $AR$; in the original KDA paper, half of the
cocoon has axial ratio of $R_{\rm T}=AR/2$. The cocoon expands along the jet axis driven
by the hotspot plasma pressure $p_{\rm h}$ and in the perpendicular direction by the cocoon
pressure $p_{\rm c}$. The rate at which energy is transported from the AGN along each jet,
i.e. the jet power $Q_{\rm jet}$, is assumed to be constant during the source's lifetime. 
The model predicts self-similar expansion of the cocoon and gives analytical formulae for 
the time evolution of various geometrical and physical parameters, e.g. the length of the
jet (Eqs.~4 and 5 in KA)

\begin{equation}
L_{\rm jet}(t)=c_{1}\left(\frac{Q_{\rm jet}}{\rho_{0}a^{\beta}_{0}}\right)^{1/(5-\beta)}
t^{3/(5-\beta)},
\end{equation}

\noindent
and the cocoon pressure (Eq.~20 in KA)

\[p_{\rm c}=\frac{18\,c_{1}^{(2-\beta)}}{(\Gamma_{\rm x}+1)(5-\beta)^{2}{\cal P}_{\rm hc}}
\left(\rho_{0}a_{0}^{\beta}\right)^{3/(5-\beta)}\times\]
\begin{equation}
\hspace{9mm}Q_{\rm jet}^{(2-\beta)/(5-\beta)}t^{-(4+\beta)/(5-\beta)},
\end{equation}

\noindent
where $c_{1}$ is a dimensionless constant (Eq.\,25 in KA), $\Gamma_{\rm x}$ is the
adiabatic index of the unshocked medium surrounding the cocoon, $t$ is the time
elapsed since the jet had started from the AGN, i.e. it is the source actual age,
and ${\cal P}_{\rm hc}\equiv p_{\rm h}/p_{\rm c}$ is the pressure ratio. In our
calculations we use the empirical formula taken from K2000

\[{\cal P}_{\rm hc}=(2.14-0.52\beta)R_{\rm T}^{2.04-0.25\beta}.\]

\subsection{Source energetics and radio power}

The energy delivered by the twin jets during the source lifetime is simply equal to
$2Q_{\rm jet}t$. Accounting for the energy losses of relativistic particles due to
adiabatic expansion within the time-dependent volume of the cocoon $V_{\rm c}(t)$,
the ratio of the jets' energy and the energy stored in the cocoon $E_{\rm tot}$ is

\[\frac{2Q_{\rm jet}t}{E_{\rm tot}}=\frac{9\Gamma_{\rm c}-4-\beta}{5-\beta}+
\frac{9(\Gamma_{\rm c}-1)}{4(5-\beta)}{\cal P}_{\rm hc},\]

\noindent
where $E_{\rm tot}(t)=u_{\rm c}(t)V_{\rm c}(t)$, $u_{\rm c}=p_{\rm c}(t)/
(\Gamma_{\rm c}-1)$ is the energy density in the cocoon, and $\Gamma_{\rm c}$ is the
adiabatic index of the cocoon's matter.

Following KDA, we assume that the cocoon is composed by many small volume elements
filled with magnetized plasma and cosmic rays which are injected into the cocoon
at the jet head --- the jet terminal shock. The initial energy distribution $n(\gamma_{\rm i})$ 
of injected particles with the Lorentz factor $\gamma_{\rm i}$ is assumed to be of a single 
power-law form with the spectral index $p$: $n(\gamma_{\rm i})=n_{0}\,\gamma_{\rm i}^{-p}$. The unknown 
normalization parameter $n_{0}$ is given by integrating the initial power-law distribution 
over the entire (assumed) energy range [$\gamma_{\rm i,min}, \gamma_{\rm i,max}$].

The relativistic particles injected into the cocoon loose their energies in a magnetic
field of energy density $u_{\rm B}$, and thus the cocoon's radio spectrum evolves with 
time. Tracing the effects of adiabatic expansion, synchrotron losses (with the assumed
effective isotropisation of the particles' pitch angle distribution), and inverse Compton 
scattering on the cosmic background radiation in the volume
elements independently, the radio power of the cocoon $P_{\nu}$ at a fixed observing
frequency $\nu$ is obtained by summing up the contributions from all volume elements,
resulting in an integral over time (see Eq.~16 in KDA). It depends on the source's age $t$
and redshift $z$, the jet power  $Q_{\rm jet}$, the half-cocoon axial ratio $R_{\rm T}$,
the exponent in the initial power-law distribution of relativistic particles $p$, the
values of the low-energy and high-energy cut-offs $\gamma_{\rm i,min}$ and $\gamma_{\rm i,max}$,
and on the ratio $\zeta$ of the magnetic field energy $u_{\rm B}$ to the energy of relativistic
electrons $u_{\rm e}$. The integral is not analytically solvable and has to be calculated 
numerically.
  
On the basis of the above analytical model we aim to predict the dynamical age and
the two physical parameters of the analyzed FR~II-type radio sources: $Q_{\rm jet}$ 
and $\rho_{0}$. In order to derive these parameters, all the other free parameters 
of the model ($\Gamma_{\rm c}$, $\Gamma_{\rm x}$, $a_{0}$,
$\beta$, $\zeta$, $p$, $\gamma_{\rm i,min}$, $\gamma_{\rm i,max}$) have to be fixed
(see Table 1).

\subsection{Selection of the free model parameters}

Following the KDA analysis, we adopt their Case~3 where both the cocoon and the ambient medium are 
assumed to be best described by the `cold' equation of state ($\Gamma_{\rm c}=\Gamma_{\rm x}=5/3$).
For the initial ratio of the energy densities of the cocoon's magnetic field and particles we 
use the equipartition condition $\zeta \equiv u_{\rm B}/u_{\rm e}=(1+p)/4$, well supported by the
X-ray observations of the lobes in powerful radio sources (see Kataoka \& Stawarz 2005, Croston 
et al. 2005, and references therein).

Selection of the particular (universal) values regarding external medium parameters $a_{0}$ 
and ${\beta}$ requires some justification. We note, that even careful 2-D modeling of a 
distribution of radio emission for powerful sources with quite regular structure can lead 
to values of $a_{0}$ discrepant with those implied by the X-ray observations of intergalactic 
thermal environment. In our simplified approach we assume $a_{0}=10$ kpc for all the analyzed 
sources, a conservative value between 2 kpc used by KDA and 50 kpc found by Wellman et al. (1997). 
We also use a constant value of $\beta =1.5$, in agreement with Daly (1995), Blundell et al. (1999), 
Willott et al. (1999), and K2000. Note, that the assumed ambient medium profile is then slightly 
flatter than the one adopted in the original KDA paper on the basis of Canizares et al.'s (1987) 
analysis, who found $\beta=1.9$ to be typical at about 100 kpc from the galactic nucleus (but see 
Gopal-Krishna \& Wiita 1987).

Another free parameter of the model is the orientation of the jet axis to the observer's
line of sight $\theta$. We assume $\theta=90\degr$ for the giant-size radio galaxies
and $\theta=70\degr$ for `normal'-size radio galaxies. This latter value is justified
on the basis of the unified scheme for extragalactic radio sources. In this scheme,
the average orientation angle for radio galaxies is $\langle\theta_{\rm RG}\rangle
\simeq 69\degr$ (Barthel 1989). The apparent size $D$ of a radio source then yields the
model cocoon length

\begin{equation}
L_{\rm jet}=D/(2\sin\theta).
\end{equation}

The next parameters which significantly influence the model predictions (as will be
shown and discussed in the next sections) are the initial power-law exponent $p$, as well 
as the low- and high-energy cut-offs of the initial electron energy distribution, 
$\gamma_{\rm i,min}$ and $\gamma_{\rm i,max}$.

The cut-off energies in the electron energy distribution injected by the jet terminal 
shock to the expanding cocoon are poorly constrained by both observations and theory. 
The values considered in the literature for $\gamma_{\rm i,max}$ vary from 10$^{4}$ up to 
$\geq 3\times 10^{8}$ (see Barai \& Wiita 2005). Here we initially study a very broad 
range $\gamma_{\rm i,max} = 10^4-10^{10}$ (section 4), in order to check how the results of 
the proposed method depend on the choice of this particular model parameter. We found 
that the effect is minor. We note in this context that the high values of the 
maximum Lorentz factors for electrons accelerated at the jet terminal shocks are more 
likely, since a number of hotspots in powerful radio sources are established sources 
of \emph{synchrotron} X-ray emission, and hence since production of high-energy 
($\gamma_{\rm i} \geq 10^7$) electrons thereby seems to be a general property of radio 
galaxies and quasars (see discussion in Hardcastle et al. 2004, Kataoka \& Stawarz 2005, 
and references therein). For these reasons, later on in the analysis (section 5) we
consider a slightly narrower range of $\gamma_{\rm i,max} = 10^7-10^{8.5}$.

As for the minimum electron energy, we note that most authors assume $\gamma_{\rm i,min} 
\rightarrow 1$, although such a choice is made largely for simplicity reasons. 
Interestingly, recently Blundell et al. (2006) argued that the X-ray observations 
of giant radio galaxy 4C~39.24 indicate an initial electron low-energy cut-off as 
high as $\gamma_{\rm i,min} \sim 10^4$ (see also in this context Carilli et al. 
1991 for the low-frequency radio observations of Cyg\,A hotspots). 
We believe, however, that these data do not show unambiguously any sharp low-energy 
cut-off of the electron energy distribution at some particular (ultrarelativistic) energy,
but instead indicate only spectral flattening occurring at low ($\gamma_{\rm i,min} < 10^4$)
electron energies. Thus, we take below $\gamma_{\rm i,min} = 1$, assuming that the particles
at the jet terminal shock are picked up by the acceleration process(es) directly from the
`thermal pool' of the cold plasma carrying the bulk of the jet kinetic energy, to form the
non-thermal electron energy distribution injected further to the expanding cocoon.
We checked however that taking $\gamma_{\rm i,min}=10$ (as favored by the analysis of
Barai \& Witta 2005) does not change our results obtained with $\gamma_{\rm i,min}=1$
presented in the next sections.

\begin{table}[h]
\caption{Observational and model parameters (see sections 2.1-2.3 for detailed descriptions).}
\begin{tabular}{lll}
\hline
       & Symbol & Dimension\\
\hline
Observational      & $z$  & [dimensionless]\\
parameters         & $D$  & [kpc]\\
derived            & $AR$ & [dimensionless]\\
from radio maps    & $\nu$& [MHz]\\
and spectrum       & $P_{\nu}$& [W\,Hz$^{-1}$sr$^{-1}$]\\
\hline
Model free         & $a_{0}$& [kpc]\\
parameters         & $\beta$& [dimensionless]\\
(to be fixed)      & $p$=2$\alpha_{\rm inj}$+1 & [dimensionless]\\
                   & $\gamma_{\rm i,min}$, $\gamma_{\rm i,max}$ &  [dimensionless]\\
                   & $\Gamma_{\rm c}$, $\Gamma_{\rm x}$ & [dimensionless]\\
                   & $\zeta$ & [dimensionless]\\
                   & $t$    & [yr]\\
                   & $\theta$ & [deg]\\
\hline
Model parameters   & $Q_{\rm jet}$ & [W]\\
derived for given  & $\rho_{0}$ & [kg\,m$^{-3}$]\\
values of $t$,     & $p_{\rm c}$& [N\,m$^{-2}$]\\
$\alpha_{\rm inj}$ and $\gamma_{\rm i,max}$  &$u_{\rm c}$ & [J\,m$^{-3}$]\\
                   & $E_{\rm tot}$ & [J]\\
\hline
\end{tabular}
\end{table}

The above choices regarding low- and high-energy electron cut-offs seem to be justified, 
although they may lead to inconsistency with the main assumption of the model, namely
with the assumed single power-law form of the initial electron energy distribution:
if the low-energy spectral flattening of the electron spectrum is the case (as has to be 
accepted if indeed $\gamma_{i, min} = 1$), it suggests that the particle distribution 
injected by the jet terminal shocks to the lobes/cocoons is characterized by some more 
complex spectral shape, for which a broken power-law could be for example a simple 
approximation. Meanwhile, in the original KDA model, universal single power-law initial 
energy distribution of the relativistic particles with the energy index $p=2.14$ was assumed, 
corresponding to the radio spectral index $\alpha_{\rm inj}=0.57$. This value agrees 
perfectly with the one obtained for the classical FR~II-type radio source Cyg\,A by means 
of fitting its radio spectrum by the `continuum injection' (C.I.) model (for the spectral 
ageing theory review see Myers \& Spangler 1985; Carilli et al. 1991). It should be 
noticed that $p=2.14$ is close to, although not exactly equal to, the model spectral 
index of relativistic particles accelerated at non-relativistic shocks by the 1st order 
Fermi (diffusive) acceleration process.

Most importantly, the low-frequency radio observations of a number of powerful radio 
sources imply a large spread of the injection spectral index, $0.38 \leq \alpha_{\rm inj} 
\leq 0.95$. This indicates that there may be no universal value of this parameter 
for all cocoons. Moreover, we note that the low-frequency radio spectra usually 
steepen with redshift (see Laing \& Peacock 1980; Macklin 1982; Gopal-Krishna 1988; 
van Breugel \& McCarthy 1990). The physical reason for such a dependence is unclear,
although the effect seems to be real. Therefore, keeping in mind all the uncertainties 
regarding $\alpha_{\rm inj}$ (or equivalently $p$) which reflect our poor understanding 
of the particle acceleration processes taking place at mildly relativistic collisionless 
astrophysical shocks, and observational limitations, we decided to treat the injection 
spectral index as a free parameter, which may be different for different sources. 

We also emphasize, that the particular value of power-law index $p$
obtained for a given source by the fitting procedure does not correspond necessarily 
to some well-defined acceleration process taking place at the jet terminal shock. Instead,
it should be considered as an \emph{effective} spectral index of the injected electron 
continuum (characterized most likely by more complex spectral shape), averaged over a very
broad energy range and over the lifetime of the source. This issue will be discussed further 
in Section 6.

\subsection{The fitting procedure}

In Table~1 we summarize all the observational and model parameters characterizing the
source's cocoon and used in the present calculations.

Assuming a value of the source age $t$, and having fixed all the free parameters of the model
as discussed in section 2.3, we find the jet power $Q_{\rm jet}$ and the central density of 
the external medium $\rho_{0}$ for an investigated source by an iterative solution of the 
system of two equations: (i) Eq.~1 equated to Eq.~3 for the jet length $L_{\rm jet}$, and 
(ii) the integral for the luminosity of the cocoon $P_{\nu}$ (Eq.~16 in KDA, see Sect.~2.2), 
requiring the matches of the solution to the observed values of $D$ and $P_{\nu}$ at a number 
of different frequencies, respectively. The above fitting procedure proved to be stable and 
always provides unique solution.

Performing the above calculations\footnote{Our programme DYNAGE to calculate the model 
parameters ($Q_{\rm jet}$, $\rho_{0}$, $p_{\rm c}$, $u_{\rm c}$, and $E_{\rm tot}$) for 
a set of given values of $t$, $\alpha_{\rm inj}$, and $\gamma_{\rm i,max}$ as described in
this paper and illustrated in Fig.~1, is available upon request from J. Machalski 
(\texttt{machalsk@oa.uj.edu.pl}).} for a number of age values $t$ assumed for a
given source, we obtain a set of solutions for $Q_{\rm jet}(t)$ and $\rho_{0}(t)$, 
which show, as expected, a correlation of $Q_{\rm jet}$ and anticorrelation of
$\rho_{0}$ with the source age $t$ for a given apparent size $D$ and total cocoon
luminosity $P_{\nu}$. An example of this effect for the four observing frequencies 
is shown in Fig.~1. A perfect 
intersection of all the $Q_{\rm jet}-\rho_{0}$ curves corresponding to different 
observing frequencies at some particular `true' age of a source is expected if 
the observed shape of its total radio spectrum (used here to determine the radio 
power at different frequencies) agrees with the theoretically predicted 
one in the framework of the C.I. model, and if the dynamical (KDA) model 
applied here is indeed correct. However, the observed spectrum of a number of the 
analyzed sources evidently departs from the predicted behavior. In such cases, the 
sources' high-frequency spectra are in particular steeper than expected, and hence 
the resultant $Q_{\rm jet}-\rho_{0}$ curve corresponding to the highest observing 
frequency bisects other curves at the age inconsistent with that implied by the 
remaining ones. 

\begin{figure}[h]
\resizebox{80mm}{!}{\includegraphics{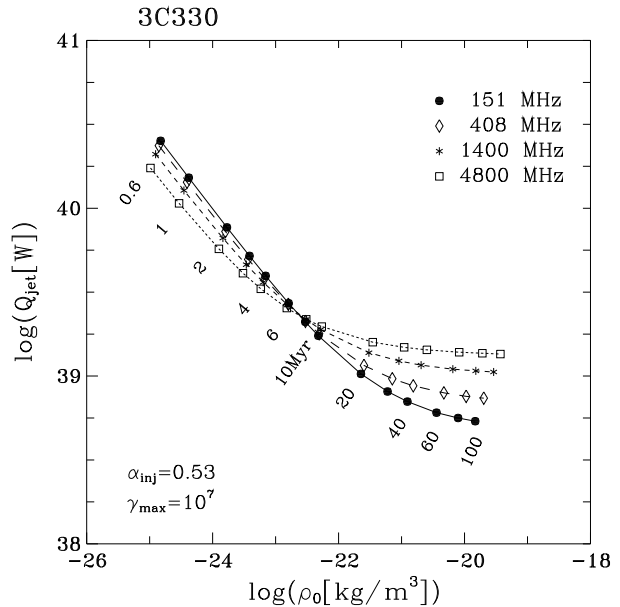}}
\caption{A set of $Q_{\rm jet}(t)$ and $\rho_{0}(t)$ solutions at the four observing
frequencies, for a number of the age values assumed for the radio galaxy 3C~330.}
\end{figure}

A `goodness' of the intersection is quantified by the $\Delta$ measure which is 
defined as follows:
indicating log\,$\rho_{0}(t)\equiv x(t)$ and log\,$Q_{\rm jet}(t)\equiv y(t)$, we calculate
the mean separation between the points determined by the values of [$x(t), y(t)$] in the
log\,$Q_{\rm jet}$--log$\rho_{0}$ plane as

\[\Delta(t)=\frac{2}{n(n-1)}\times\]
\begin{equation}
\times \sum_{i=1}^{n-1}\left(\sum_{j=i}^{n-1}\left\{[x_{i}(t)-x_{j+1}(t)]^{2}+
[y_{i}(t)-y_{j+1}(t)]^{2}\right\}^{1/2}\right),
\end{equation}
\noindent
where $n$ is the number of the observing frequencies included. A minimum of $\Delta(t)$ 
(i.e. `the best' intersection of the $Q_{\rm jet}-\rho_{0}$ curves for different
frequencies) is considered as an indicator of the source's real dynamical age $t$. This
minimum also discriminates a unique values of $Q_{\rm jet}$ and $\rho_{0}$, as well as
$p_{\rm c}$ and $E_{\rm tot}$ for the considered source. Hereafter, the age corresponding 
to such a set of $Q_{\rm jet}$ and $\rho_{0}$ values is called the {\sl age solution}.

\begin{table*}[t]
\center
\caption{List of sources and their observational parameters derived from radio maps
and spectra. The radio power is in units of W Hz$^{-1}$ sr$^{-1}$}
\begin{tabular*}{130mm}{llrrrcccc}
\hline
Source name & z & D[kpc] & AR & $\theta[\degr]$ &log\,$P_{151}$ & log\,$P_{408}$ &
log\,$P_{1400}$ & log\,$P_{4800}$\\
\hline
Cyg\,A & 0.0564 & 135  & 3.9 & 80 & 27.796 & 27.465 & 26.963 & 26.358\\
3C55   & 0.7348 & 496  & 6.4 & 70 & 27.644 & 27.247 & 26.718 & 26.148\\
3C103  & 0.330  & 458  & 7.0 & 70 & 26.906 & 26.630 & 26.140 & 25.552\\
3C165  & 0.296  & 384  & 3.4 & 70 & 26.579 & 26.244 & 25.769 & 25.227\\
3C239  & 1.786  &  96  & 2.7 & 70 & 28.483 & 28.092 & 27.541 & 26.914\\
3C247  & 0.749  &  95  & 4.8 & 70 & 27.350 & 27.117 & 26.746 & 26.284\\
3C280  & 0.996  & 113  & 2.9 & 70 & 27.990 & 27.697 & 27.288 & 26.829\\
3C292  & 0.71   & 960  & 8.4 & 90 & 27.406 & 27.020 & 26.547 & 26.075\\
3C294  & 1.779  & 135  & 3.8 & 70 & 28.412 & 27.984 & 27.431 & 26.841\\
3C322  & 1.681  & 283  & 5.0 & 70 & 28.171 & 27.906 & 27.485 & 26.954\\
3C330  & 0.549  & 395 & 10.9 & 90 & 27.432 & 27.150 & 26.771 & 26.334\\
3C332  & 0.1515 & 229  & 4.9 & 70 & 25.744 & 25.460 & 25.068 & 24.631\\
B0908+376 & 0.1047&  72 & 2.3 & 70 & 24.694 & 24.452 & 24.120 & 23.755\\
B1209+745 & 0.107 & 795 & 5.8 & 90 & 24.919 & 24.583 & 24.152 & 23.706\\
B1312+698 & 0.106 & 794 & 5.5 & 90 & 25.174 & 24.883 & 24.497 & 24.066\\
\hline
\end{tabular*}
\end{table*}

However, the calculations indicate that the `age solution' significantly depends on
the exponent in the initial power-law energy distribution, $p$. In addition, they suggest 
that there is no canonical value of this injection spectral index which would be good for 
each and every source. (This in fact agrees with the preliminary notes given at the end of 
the previous subsection 2.3; this issue will be discussed further in section 6.)
For this reason, the best value of $p$ (or equivalently $\alpha_{\rm inj}$), needed for a
precise estimate of the dynamical age of a particular FR~II-type radio source, is determined 
below by finding the minimum of the jet kinetic energy delivered to the cocoon (see section 4.3). 
The age solution corresponding to the minimum jet energy is called hereafter the {\sl best 
solution}. 

\section{The data}

The data necessary for the age determination of a given radio source with the method
presented above are: its projected linear length $D$, its axial ratio $AR$, and its total 
radio power $P_{\nu}$ at a number of observing frequencies, and the source's redshift, $z$.

As we limit our method to FR~II-type sources only, $D$ is invariant of frequency and can
be precisely determined from the radio maps. The lateral size (base diameter $b$ of the
cocoon) is taken as the average of the widths of the two lobes. They are determined on the 
maps with the highest sensitivity to surface brightness as the deconvolved width of a 
transversal cross-section through the lobes measured half-way between the core and the 
outer edge of the source. Thus, $AR=D/b$.

The source power $P_{\nu}$ is determined at four frequencies: $\nu$=151, 408, 1400, and
4800 MHz. Because not all sources were observed at these particular frequencies,
the cocoon's spectrum has been derived from the available flux densities with the
contributions from the radio core and the hotspots subtracted if possible. Next,
the simple analytical functions: $y=a+bx+c\,\exp(\pm x)$, where $x=\log\nu$[GHz] and
$y=\log\,S(\nu)$[mJy], have been fitted to these flux densities weighted by  their given
error. These two functions were found to provide the best fit (the least $\chi^{2}$) to
the data. Then, the values of $P_{151}$, $P_{408}$, $P_{1400}$ and $P_{4800}$ have been
calculated using flux densities found with the above functions and the appropriate radio
K-corrections\footnote{There is no need to use the same frequencies for different sources;
a few observing frequencies should cover a wide range of the radio spectrum}.   

The FR~II-type radio sources analyzed in this paper, and their observational parameters
as used in the fitting procedure, are given in Table~2. These sources, subjectively selected
from the sample of Machalski, Chy\.{z}y \& Jamrozy (2004a), include high-redshift and
low-redshift, high-luminosity and low-luminosity, giant-sized and normal-sized radio
galaxies. The sources in Table~2 do not include quasars whose
jets' axis (and hence radio structure) may be strongly projected onto the sky.   

\section{Fitting result for Cyg\,A }

In this section the dynamical age estimate and its dependence on the values of 
the model free parameters is presented for Cyg\,A.

\subsection{The age solution and its dependence on $\alpha_{\rm inj}$ and $\gamma_{\rm i,max}$}

Assuming $\theta=80\degr$ and adopting (after K2000) parameters $p=2.14$ and $\gamma_{\rm i,max}=10^{4}$,
the fitted values of $Q_{\rm jet}$ and $\rho_{0}$ for different hypothetical ages of
this source are shown as the log\,$Q_{\rm jet}$--log\,$\rho_{0}$ diagram in Fig.~2. This figure
shows that the four curves corresponding to the four observing frequencies intersect 
at the age of about 10 Myr. The age solution giving a minimum of the $\Delta(t)$ measure for the 
set of the model free parameters (recalled in the caption of Fig.~2) is $t=10.9\pm 0.9$ Myr.

\begin{figure}[h]
\resizebox{\hsize}{!}{\includegraphics{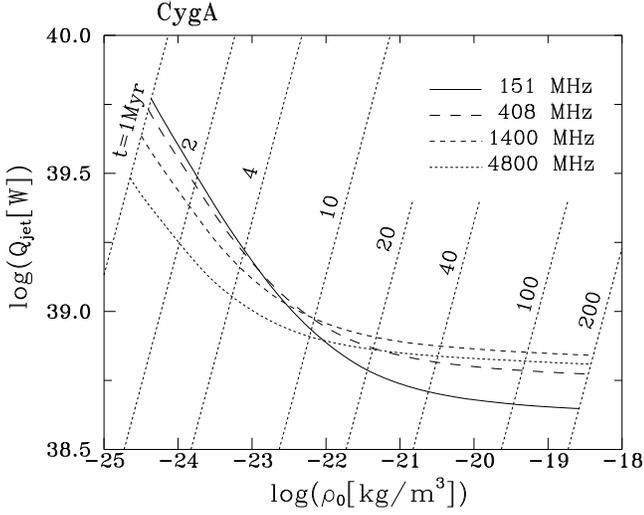}}
\caption{$Q_{\rm jet}-\rho_{0}$ diagram for Cyg\,A at the four observing frequencies.
The dotted diagonal lines indicate an age related to the fitted $Q_{\rm jet}$ and
$\rho_{0}$ values. Other free parameters for all the curves are: $R_{\rm T}$=1.95,
$\beta$=1.5, $\alpha_{\rm inj}$=0.57, and $\gamma_{\rm i,max}$=10$^{4}$ (see also section 2.3)}
\end{figure}

As mentioned previously, a minimum of $\Delta(t)$ appears to be dependent on the assumed 
parameters of the initial energy distribution of relativistic particles. The dependence 
of $\Delta$ on $t$ for two different (rather extreme) values of $\gamma_{\rm i,max}=10^4$ 
and $=10^{10}$ is plotted in Fig.\,3 for different values of $\alpha_{\rm inj}$. As shown, 
the age solution is not very sensitive for the particular choice of $\gamma_{\rm i,max}$,
especially if $\alpha_{\rm inj} > 0.5$.

The dependence of the resulting age solution on $\alpha_{\rm inj}$ for the same 
two extreme values of $\gamma_{\rm i,max}$ is shown in Fig.\,4. It is 
clear that the age solution strongly increases with the decreasing 
$\alpha_{\rm inj}$, and hence that the minimum $\Delta$ measure is smaller for 
larger assumed values of $\alpha_{\rm inj}$. This is due to the fact that in the 
case of flatter and flatter initial electron energy distributions more and more 
time has to elapse for the radio continuum to steepen up to some given (observed) 
level. As a result, the obtained age solution anticorrelates with the injection 
spectral index, and hence also the minimum of $\Delta$ decreases with decreasing $t$. 
Such a decrease continues until reaching the point where the radio continuum did not 
have time to steepen at all, correspondingly to the situation when all the curves 
on the $Q_{\rm jet} - \rho_0$ overlay each other. Such an ambiguous situation leads 
to the question which value of $\alpha_{\rm inj}$ one should really assume for a 
given source to apply the presented method of the dynamical age estimate. This issue 
will be addressed further in section 4.3 below.

\begin{figure}[t]
\resizebox{76mm}{!}{\includegraphics{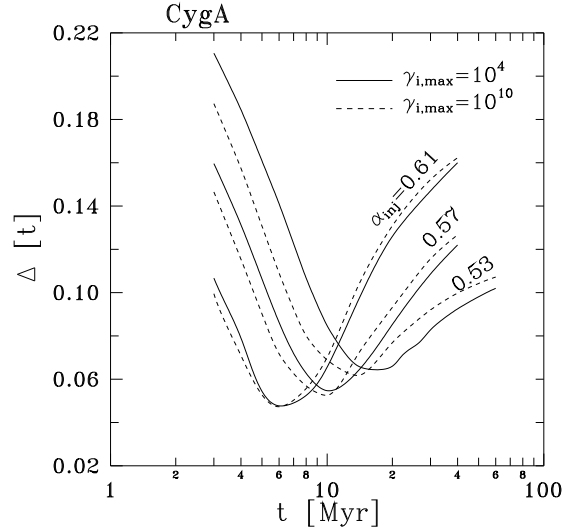}}
\caption{The influence of the initial energy distribution parameters on the `goodness'
of fit quantified by the $\Delta$ measure for Cyg\,A.}
\end{figure}

\begin{figure}[t]
\resizebox{76mm}{!}{\includegraphics{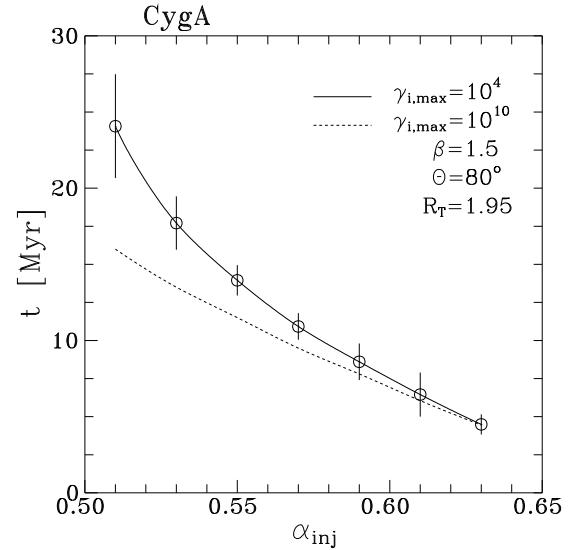}}
\caption{The influence of $\alpha_{\rm inj}$ on the age estimate for Cyg\,A, with a set
of the fixed model parameters.}
\end{figure}

\begin{figure}[h]
\resizebox{\hsize}{!}{\includegraphics{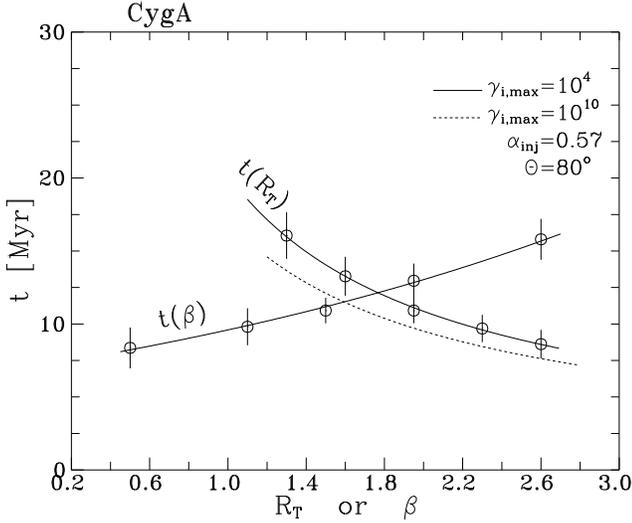}}
\caption{The influence of $\beta$ and  $R_{\rm T}$ on the age estimate for Cyg\,A, with a
set of the fixed model parameters.}
\end{figure}

\subsection{Dependence on $\beta$, $AR$, and $\theta$}

The fitting results indicate that the effect due to changing particular values of the 
$\beta$, $AR$ and $\theta$ parameters is of minor importance for the resulting age solution.
Fig.\,5 shows the influence of the exponent $\beta$ in the density distribution of the 
ambient medium and of the source axial ratio $AR\equiv R_{\rm T}\times 2$ on the age 
estimate. (Though the $AR$ parameter is the observed one, it is much less precisely
measured than the source linear size $D$.). The influence of the unknown inclination angle
$\theta$ on the age is also small. Comparison of Eqs.\,(1) and (3) yields
$t\propto(D/\sin\theta)^{(5-\beta)/3}$. Thus, $t$ increases no more than 1.21 and 1.15
for $\beta$=1 and $\beta$=2, respectively, if $\theta$ decreases from 90$\degr$ to
60$\degr$. We conclude that the influence of $\theta$ on $t$ is comparable to that
caused by the uncertainty in the $AR$ determination.

\subsection{Dependence of the age solution on kinetic energy delivered by the jet}

As noted in section 4.1, the fitted jet power increases with decreasing age (and hence
with increasing $\alpha_{\rm inj}$) for a given
size and brightness of the source (see Figs.\,1 and 2). In Fig.\,6 we show both the fitted 
jet power and the age of Cyg\,A as a function of $\alpha_{\rm inj}$ for the two extreme 
values of $\gamma_{\rm i,max}$ considered in this section. As the parameters $Q_{\rm jet}$
and $t$ depend on $\alpha_{\rm inj}$ in the opposite way, there is always a minimum value
of the product $Q_{\rm jet}\times t$. This product gives the kinetic energy delivered to
the cocoon by the jet. The fitting procedure indicates thus that this total
energy has a minimum for some specific value of $\alpha_{\rm inj}$ and fixed observational 
parameters.

\begin{figure}[t]
\resizebox{\hsize}{!}{\includegraphics{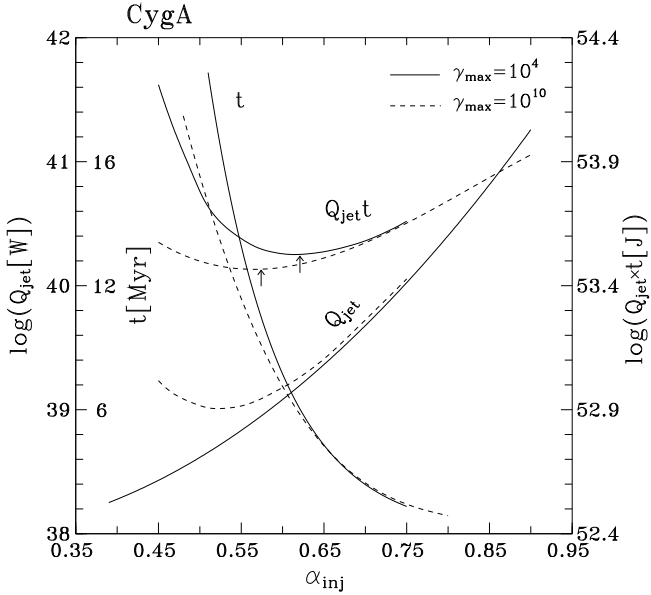}}
\caption{Diagram of the jet power $Q_{\rm jet}$, the age $t$, and the total energy 
$Q_{\rm jet}\,t$ delivered by the Cyg\,A jet to its cocoon, as functions of 
$\alpha_{\rm inj}$ for the two (extreme) values of $\gamma_{\rm i,max}$. 
The arrows indicate $\alpha_{\rm inj}$ values corresponding to the minimum 
total cocoon energy.}
\end{figure} 

Basing on the above effect, we consider the age corresponding to the minimum kinetic energy
as the `best solution' for the dynamical age of a given source (see section 2.4).
The `best solution' for Cyg\,A gives $t$=$10.4\pm1.6$ Myr ($\alpha_{\rm inj}$=$0.574\pm0.011$)
and $t$=$5.7\pm 0.4$ Myr ($\alpha_{\rm inj}$=$0.621\pm 0.007$) for $\gamma_{\rm i,max}$=$10^{4}$
and $\gamma_{\rm i,max}$=$10^{10}$, respectively. Since one should expect in reality
$\gamma_{\rm i,max}$ in between of the two extreme values considered here for illustrative purposes,
we safely conclude that the presented method indicates the age and the injection spectral
index of Cyg\,A to be $t=6-10$ Myr and $\alpha_{\rm inj} = 0.57-0.62$, respectively,
in a very good agreement with other dynamical and spectral estimates presented in the literature
(see, e.g., Kino \& Kawakatu 2005). In addition, the corresponding kinetic power of the Cyg\,A
jet is $Q_{\rm jet} = (1-2) \times 10^{39}$ W, which is a factor of $2-3$ higher than the value
estimated by K2000, and a factor of $2$ lower than the one obtained by Kino \& Takahara (2004)
by means of detailed analysis of the Cyg\,A hotspots' radiative properties.

\section{Fitting results for other radio galaxies}

\subsection{The `best solution' for the age}

Further results of the age estimation derived with our method are presented in more details
for two other sources: giant radio galaxy B\,1312+698 and normal-sized galaxy 3C\,55.
The appropriate $Q_{\rm jet}$ vs. $\rho_{0}$ diagrams for the given values of $\alpha_{\rm inj}$ 
and $\gamma_{\rm i,max}$ are shown in Figs.\,7a and 8a. As previously, the dotted
diagonal lines indicate an age related to the values of $Q_{\rm jet}$ and $\rho_{0}$ fitted
to observations at the four different frequencies. The ideal, in this case, intersections of
the four curves gives the precise `age solutions', as well as the precise estimates of the 
jet powers and the central core densities for both sources.

\begin{figure*}[t]
\resizebox{176mm}{!}{\includegraphics{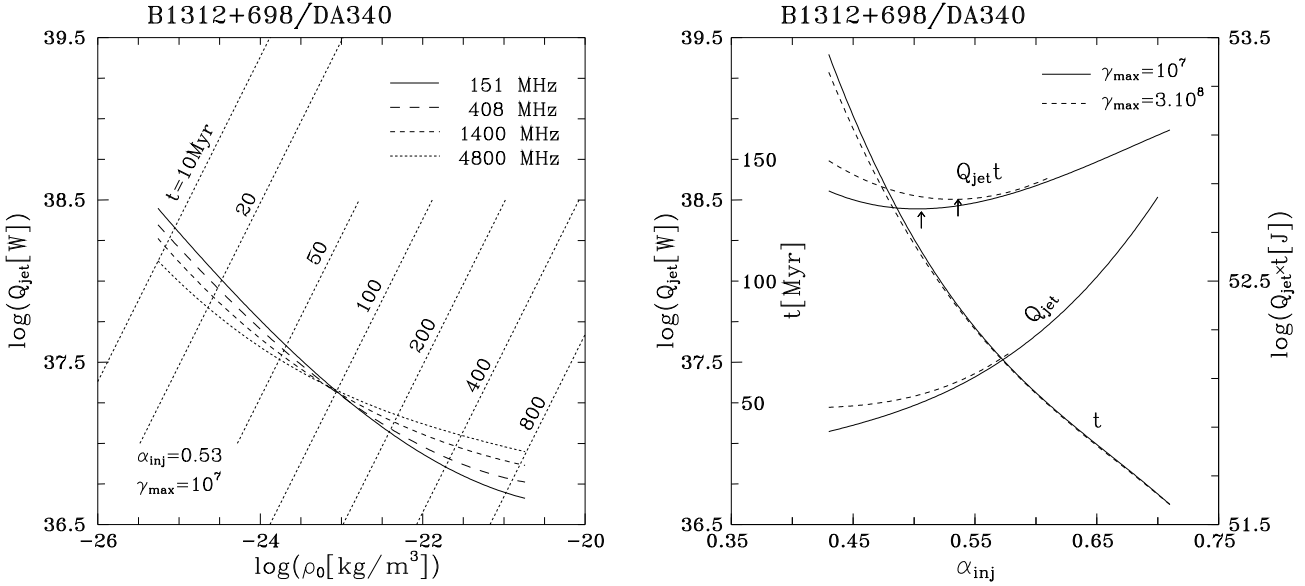}}
\caption{{\bf a)} $Q_{\rm jet}-\rho_{0}$ diagram for the giant radio galaxy
B\,1312+698: the `age solution' for $\alpha_{\rm inj}$=0.53 and $\gamma_{\rm i,max}$=10$^{7}$.
{\bf b)} $Q_{\rm jet}$, $t$, and $Q_{\rm jet}\times t$ vs. $\alpha_{\rm inj}$ diagrams for
two different values of $\gamma_{\rm i,max}$. The minima of the total energy corresponding to
the `best solution' are marked with arrows.}
\end{figure*}

The dependence of these `age solutions' (marked as $t$ for simplicity) on $\alpha_{\rm inj}$
and $\gamma_{\rm i,max}$ is presented in Figs.\,7b and 8b, together with the $Q_{\rm jet}-
\alpha_{\rm inj}$ and $Q_{\rm jet}\, t-\alpha_{\rm inj}$ curves. Clearly, the `best solutions'
exist within the theoretically acceptable range of $0.5 \leq \alpha_{\rm inj} \leq 0.7$. 

The `best solutions' for the dynamical age of all the sources from Table~2 with the
corresponding values of $\alpha_{\rm inj}$ are collected in Table~3. In order to show
how these solutions depend on the assumed value of the maximum electron energies, we
give the best solution for $\gamma_{\rm i,max}=10^{7}$ and $3 \times 10^{8}$. As shown, the
differences are minor. In particular, column 6 of Table~3 gives the ratio of the standard
deviation to the mean of the ages derived with these two different values $\gamma_{\rm i,max}$,
showing that the resulting age estimates differ from $\sim$4\% to $\sim$30\%, mostly by
$\sim$10\% only.

\begin{figure*}[t]
\resizebox{176mm}{!}{\includegraphics{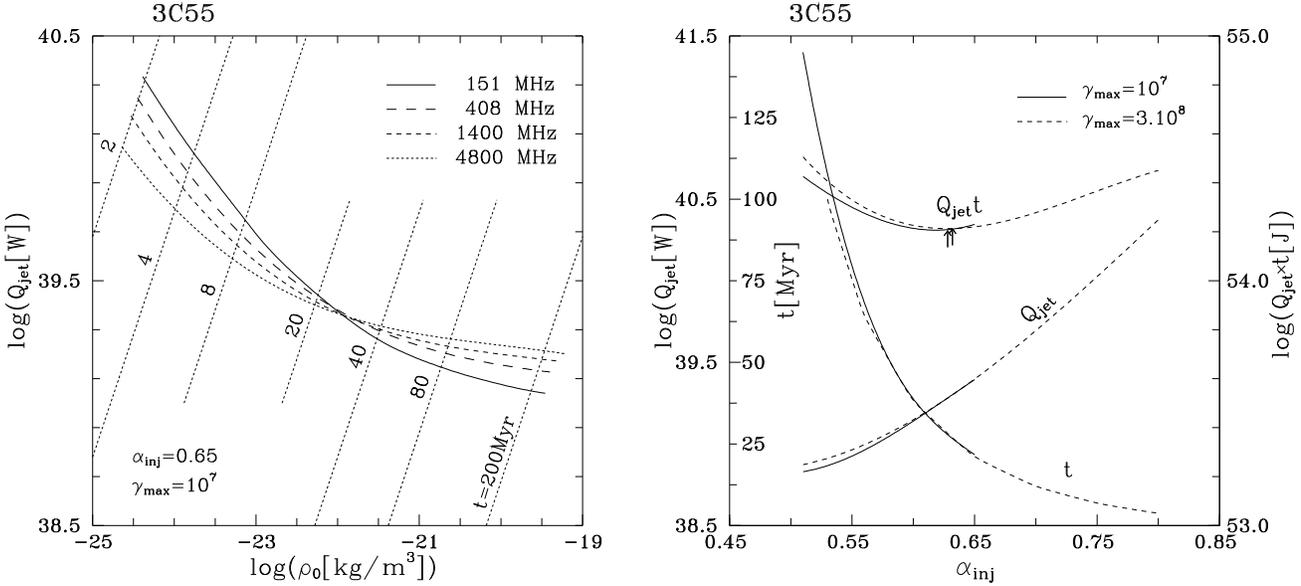}}
\caption{The same as in Fig.\,7 but for the radio galaxy 3C\,55.}
\end{figure*}

\begin{table*}[th]
\center
\caption{$\alpha_{\rm inj}$ values relating to the minimum total energy of the jet for two
different values of $\gamma_{\rm i,max}$, and the
corresponding `best solution' of age for the analyzed sample of FR~II-type radio galaxies.
Column 6 gives the ratio of the standard deviation to the mean of the ages in columns
3 and 5. Columns 7 and 8 give $\alpha_{\rm inj}$ value fitted for the C.I. model using
the {\sc SYNAGE} package of Murgia (1996), and the synchrotron (radiative) age of the sources
compiled from the literature, respectively}
\begin{tabular*}{148mm}{lcccccccc}
\hline
  &   $\gamma_{\rm max}$=10$^{7}$ &&
    $\gamma_{\rm max}$=3 $\times$ 10$^{8}$  \\
Source & $\alpha_{\rm inj}$ & $t$[Myr] & $\alpha_{\rm inj}$ & $t$[Myr] & $\Delta t/\overline{t}$
 & $\alpha_{\rm CI}$ & $\tau_{\rm rad}$[Myr] & $t/\tau_{\rm rad}$\\
1 & 2 & 3 & 4 & 5 & 6 & 7 & 8 & 9 \\
\hline
Cyg\,A  & 0.607 & 6.78 & 0.618 & 5.94 & 0.09 & 0.571$\pm$.023 & 6.4$\pm$1.4 & 1.01\\
3C55    & 0.628 & 26.7 & 0.632 & 24.1 & 0.07 & 0.630$\pm$.060 & 9.4$\pm$1.6 & 2.84\\
3C103   & 0.576 & 42.8 & 0.589 & 36.5 & 0.11 & 0.617$\pm$.099 &10.6$\pm$1.4 & 4.04\\
3C165   & 0.557 & 80.5 & 0.580 & 64.5 & 0.16 & 0.523$\pm$.051 & 41$\pm$5    & 1.96\\
3C239   & 0.673 & 2.88 & 0.678 & 2.72 & 0.04 & 0.716$\pm$.033 &2.34$\pm$0.15& 1.23\\
3C247   & 0.546 & 1.67 & 0.570 & 1.31 & 0.17 & 0.599$\pm$.021 &1.35$\pm$0.15& 1.24\\
3C280   & 0.560 & 2.61 & 0.575 & 2.22 & 0.11 & 0.694$\pm$.031 &2.52$\pm$0.2 & 1.04\\
3C292   & 0.566 & 38.1 & 0.577 & 36.0 & 0.04 & 0.782$\pm$.049 & 8.4$\pm$2   & 4.54\\
3C294   & 0.683 & 3.65 & 0.686 & 3.48 & 0.03 & 0.898$\pm$.067 &2.85$\pm$0.1 & 1.28\\
3C322   & 0.557 & 4.00 & 0.569 & 3.93 & 0.01 & 0.624$\pm$.041 &3.65$\pm$0.3 & 1.10\\
3C330   & 0.543 & 6.55 & 0.563 & 5.26 & 0.15 & 0.580$\pm$.039 &10.1$\pm$1.5 & 0.65\\
3C332   & 0.528 & 35.7 & 0.555 & 26.1 & 0.22 & 0.620$\pm$.040 & 23.5$\pm$4.5& 1.52 \\
B0908+376 & 0.514 & 19.1 & 0.543 & 12.6 & 0.29 &0.548$\pm$.082 & 14$\pm$2.5 & 1.36\\
B1209+745 & 0.512 & 151.3& 0.539 & 128.8& 0.11 &0.764$\pm$.042 & 81$\pm$15  & 1.87\\
B1312+698 & 0.506 & 112.3& 0.536 & 88.8 & 0.17 &0.639$\pm$.035 & 41$\pm$8   & 2.74\\

\hline
\end{tabular*}
\end{table*}

\subsection{Comparison of the `best solution' dynamical ages with the synchrotron ages}

A quality of the dynamical age estimated with the method presented here can be justified by
its comparison with the synchrotron (radiative) age, $\tau_{\rm rad},$ derived from the spectral
ageing analysis (e.g. Myers \& Spangler 1985; Carilli et al. 1991). In this analysis the radiative
age of emitting particles is determined from the `break' frequency in the observed radio
spectrum and the magnetic field strength, usually computed under the equipartition conditions.
In order to find this break frequency, the $\alpha_{\rm inj}$ value must be known. In most
of the published papers concerning the above analysis, the value of $\alpha_{\rm inj}$ was
usually identified with the slope of the observed low-frequency spectrum. Because of a
reasonable criticism of this simple method (see section 1), Murgia (1996) developed the
software {\sc SYNAGE} which allows the best fit of the spectral data to the theoretical
models of the energy losses by constraining the values of the most important model parameters,
especially a value of $\alpha_{\rm inj}$.

The $\alpha_{\rm inj}$ value for the C.I. model, $\alpha_{\rm CI}$, fitted to the spectrum
of the lobes of the sources from Table~2 using the {\sc SYNAGE} algorithm, is given in
column 7 of Table~3. It is easy to see that the values of $\alpha_{\rm CI}$ are typically
higher than the values of $\alpha_{\rm inj}$ determined by minimizing the total energy as
proposed in this paper, though the errors in $\alpha_{\rm inj}$ are of the same order as
the errors in $\alpha_{\rm CI}$ (also given in column 7). Only for Cyg\,A and (formally)
for 3C165 the fitted values of $\alpha_{\rm CI}$ are lower than the values of $\alpha_{\rm inj}$.

A discrepancy between the values of $\alpha_{\rm CI}$ and $\alpha_{\rm inj}$ for the other
sources can be explained either by (i) a lack of flux density data at frequencies below
$\sim$100 MHz (which may be crucial for a proper determination of the slope of the initial spectrum)
and the uncertainties of the available data, or by (ii) an influence  of other physical processes
not taken into account in the KDA model, such as reacceleration of the relativistic particles,
non-standard evolution of the cocoon's magnetic fields, etc. Note, that the $\alpha_{\rm CI}$
values, fitted with the {\sc SYNAGE} and being much higher than the effective values of
$\alpha_{\rm inj}$ determined by the minimum jet energy delivered to the source (cocoon), may lead
to a most unlikely young age and a corresponding enormously high jet-head advance velocity, $v_{\rm h}$.
For example, $\alpha_{\rm inj}\approx \alpha_{\rm CI}$=0.898 for the high-redshift radio galaxy
3C294 (see Table~3) would yield the dynamical age of 0.25 Myr only, which would imply the
unacceptable high value of $v_{\rm h}$=0.936\,$c$!

We note in this context, that large ($\sim 0.3\, c$) advance velocities are, in fact, observed, 
however only in `compact symmetric objects' (CSOs; see, e.g., Owsianik \& Conway 1998, 
Polatidis \& Conway 2003, Gugliucci et al. 2005). These sources, characterized by the 
projected linear sizes less than $\sim$1 kpc, are believed to be young versions of the 
extended radio galaxies, possibly even of the powerful classical doubles (see a review by O'Dea 
1998). And indeed, the implied kinetic ages (between $100$ yr and $10^4$ yr) usually agree
well with the spectral ages determined for some of these objects assuming energy equipartition 
within the lobes (Katz-Stone \& Rudnick 1997). On the other hand, it 
should not be surprising that the advanced velocities of the jets in CSOs are larger than the 
ones in `regular' FR~II radio galaxies, for which Scheuer (1995) gives an upper limit of 
$0.1 \, c$. That is because of the differences in the lobes/jets evolution for compact and 
extended objects, caused for example by the differences in the ambient medium density profiles. 
In particular, Kawakatu \& Kino (2006) argued for constant jet advanced velocity of evolving 
young/compact radio sources, while for the decreasing advanced velocity with increasing linear
size at later evolutionary stages (see also in this context Alexander 2000). Therefore, here 
we take $0.3\, c$ as an upper limit for the `realistic' values of the jet advanced velocities
in all the FR~II radio galaxies considered in this paper, in agreement with Scheuer (1995), and
multi-epoch observations of young radio sources.

Finally, the synchrotron ages $\tau_{\rm rad}$ (with associated errors) for all
the sources in Table~2 ---
collected from the papers of Alexander \& Leahy (1987), Leahy et al. (1989), Liu et al. (1992),
and Parma et al. (1999) --- are given in column 8 of Table~3. As shown, the synchrotron ages and
our dynamical age estimates are comparable for the young sources, i.e. those whose ages are less
or much less than 10 Myr. Such behavior is in fact consistent with the conclusions of
Blundell \& Rawlings (2000), who argued that the spectral ageing analysis is most likely erroneous
for lobes older than 10 Myr. Indeed, in the case of the largest (and oldest) sources from
Table~3 (3C~55, 3C~103, 3C~292, B1209+745, B1312+698), the ratios of the dynamical age estimates to
the synchrotron ages (see column 9 of Table 3) rise to a factor of $2-4$. This agrees also with
the earlier result on this ratio determined for the three giant radio galaxies: J0912+3510,
J1343+3758, and J1451+3357 (Jamrozy \& Machalski 2005). The `best solution' for the age of the
analyzed radio galaxies vs. their synchrotron age is plotted in Fig.\,9.


\begin{figure}[h]
\resizebox{\hsize}{!}{\includegraphics{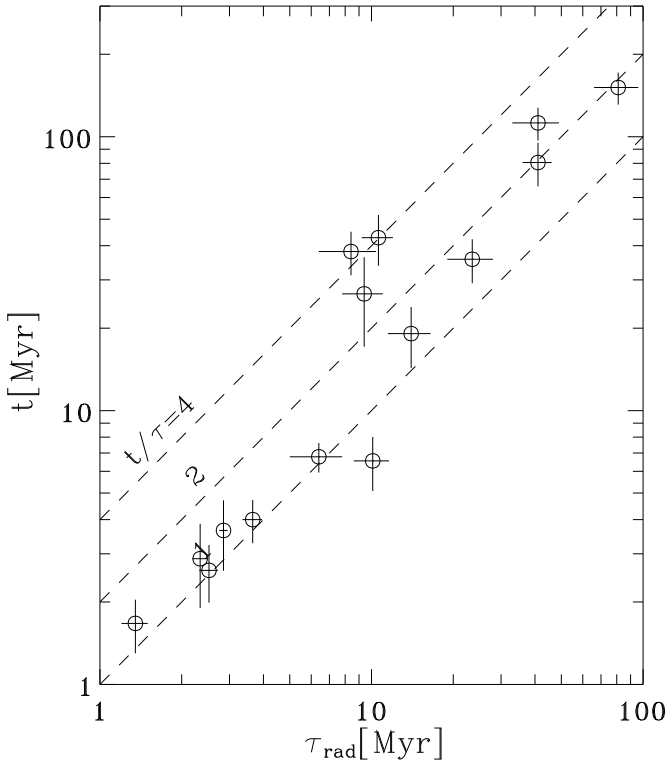}}
\caption{`Best solution' age vs. the synchrotron age for the radio galaxies from Table 2.}
\end{figure}

\subsection{The jet power, the central core density, and other physical parameters}                                                                                                                                                                                                                                                                  

Given the age determined for a given source, we can derive its other physical parameters.
It was shown in section 4.1 that each `age solution' determined by the intersection
of the age curves in the $Q_{\rm jet}-\rho_{0}$ diagrams (Figs.\,2, 7a, and 8a) indicates
unique values of the jet power $Q_{\rm jet}$ and the central core density $\rho_{0}$. It
also provides unique values of the source (cocoon) energy density and pressure, $u_{\rm c}$
and $p_{\rm c}$, respectively, as well as the total radiative energy $E_{\rm tot}$ (see
Table~1). We note, that the minimum total cocoon energy (which gives the `best solution'
of the age) naturally corresponds to the equivalent minimum of $u_{\rm c}$ and $p_{\rm c}$.

\begin{table*}[ht]
\center
\caption{Physical parameters of the sources from Table 2}
\begin{tabular*}{168mm}{llllllll}
\hline
Source & log($Q_{\rm jet}$[W]) & log($\rho_{0}$[kg\,m$^{-3}$]) & log($u_{\rm c}$[J\,m$^{-3}$])
& log($p_{\rm c}$[N\,m$^{-2}$]) & log($E_{\rm tot}$[J]) & $Q_{\rm jet}t/E_{\rm tot}$&$v_{\rm h}/c$\\
&\\
\hline
Cyg\,A      & 39.19 & $-$22.45 & $-$10.42 & $-$10.59 & 53.17 & 2.24 & 0.032\\
3C55        & 39.28 & $-$22.02 & $-$11.24 & $-$11.42 & 53.68 & 3.38 & 0.030\\
3C103       & 38.54 & $-$21.92 & $-$11.63 & $-$11.81 & 53.10 & 3.70 & 0.017\\
3C165       & 37.99 & $-$22.46 & $-$12.06 & $-$12.22 & 53.08 & 2.06 & 0.008\\
3C239       & 39.90 & $-$22.91 & $\;\,-$9.93 & $-$10.11 & 53.60 & 1.84 & 0.054\\
3C247       & 39.19 & $-$23.75 & $-$10.50 & $-$10.68 & 52.51 & 2.61 & 0.093\\
3C280       & 39.52 & $-$23.50 & $-$10.51 & $-$10.68 & 53.17 & 1.90 & 0.071\\
3C292       & 39.18 & $-$22.29 & $-$11.83 & $-$12.01 & 53.63 & 4.54 & 0.082\\
3C294       & 39.95 & $-$22.65 & $-$10.01 & $-$10.18 & 53.67 & 2.21 & 0.060\\
3C322       & 39.91 & $-$23.30 & $-$10.82 & $-$11.00 & 53.58 & 2.70 & 0.115\\
3C330       & 39.41 & $-$22.71 & $-$11.15 & $-$11.33 & 52.93 & 6.26 & 0.098\\
3C332       & 37.45 & $-$22.61 & $-$12.06 & $-$12.24 & 52.08 & 2.67 & 0.010\\
B0908+376   & 36.52 & $-$23.62 & $-$12.23 & $-$12.41 & 51.06 & 1.73 & 0.006\\
B1209+745   & 36.98 & $-$22.80 & $-$13.37 & $-$13.54 & 52.17 & 3.07 & 0.009\\
B1312+698   & 37.20 & $-$23.00 & $-$13.25 & $-$13.42 & 52.33 & 2.93 & 0.012\\
\hline
\end{tabular*}
\end{table*}

All the physical parameters corresponding to the `best solution' of age for the sources
considered in this paper, obtained with $\gamma_{\rm i,max}$=10$^{7}$ and given in
Table~3, are collected in Table~4. The last column of Table~4 gives in addition the ratio 
of the mean expansion velocity of the jet head to the speed of light, i.e. one half of 
the projected linear size of a given source divided by its age estimate.

\subsection{A limitation of the method}

A justified question may be risen: starting from what age/size (or equivalently stage of
the source's evolution) the described method can be applied? \emph{A priori} the method
is applicable to the sources for which radio emission is dominated by the 
lobes or cocoons evolving  within a power-law ambient medium 
density profile. 
Therefore, our method cannot be applied directly to the young (CSO-type) sources whose 
linear sizes are $< 1$ kpc, because it is unlikely that  
their observed radio emission may be attributed 
entirely to the non-relativistic lobes. In particular, in many of such sources contribution
from the underlying relativistic jets and bright mildly-relativistic hotspots to the
total radiative output may be significant. Finally, it is also not obvious if some (small)
fraction of CSOs is `frustrated' by the dense ambient medium, as advocated by many 
authors (see, e.g., De Young 1993, Alexander 2000).

Slightly larger versions of CSOs, called `medium symmetric objects' (MSOs, or `compact 
steep-spectrum' sources, CSS, if the spectral characteristics are considered), are
probably more promising in this context, since their projected linear sizes are
typically $1-10$ kpc. For example, the dynamical age estimates determined
with our method for two of such objects from the sample of Murgia et al. (1999) are
$142\pm 35$ kyr ($\alpha_{\rm inj}$=0.571$\pm$0.020) for 3C~67 with $D=11$ kpc, and
$258\pm 67$ kyr ($\alpha_{\rm inj}$=0.753$\pm$0.100) for 3C~186 with $D=15.7$ kpc.
The above dynamical age estimates are larger by a factor of $2-3$ than the synchrotron 
ages found by Murgia et al. (which are $51$ and $113$ kyr for 3C~67 and 3C~186, 
respectively). However, such a discrepancy may be expected, as these synchrotron 
ages were determined using the break frequency in the spectrum of the entire source. 
That spectrum must be dominated by the radiation produced by the hottest regions of 
a given object, likely very close to a jet terminal shock. For this reason, the 
spectral ages found by Murgia et al. can be significantly lower than the dynamical ages
derived by us.
On the other hand, it is worth to emphasize that the $\alpha_{\rm inj}$ values fitted by
Murgia et al. in the frame of the C.I. model and our values determined from the
minimum jet kinetic energy  agree between themselves very well.

\section{Remarks on the injection electron spectrum}

Jets in powerful radio sources were considered as being characterized by sub- or
eventually mildly-relativistic bulk velocities on large (hundreds of kpc) scales,
$v_{\rm jet} < 0.95 \, c$ (e.g., Wardle \& Aarons 1997). Also, advanced velocities
of their terminal features, the hotspots, were known to be non-relativistic,
$v_{\rm hs} < 0.1 \, c$ (Scheuer 1995). For these reasons, it was widely assumed
that the double shock structures formed at the heads of the jets in quasars and FR~II
radio galaxies were in fact always non-relativistic phenomena. In particular, it was
argued that the non-relativistic reverse shock formed at the jet terminal point was
the place of the efficient diffusive (Fermi I) particle acceleration process, leading
to formation of the power-law particle energy spectrum within the hotspot, injected later
to the expanding lobes/cocoons by means of the plasma backflow.

It is known that the well-established theory for particle shock acceleration in the
non-relativistic test-particle limit predicts a universal power-law index of the 
accelerated particle continuum $p = 2$, weakly dependent on the plasma conditions
on both sides of the shock (e.g., Blandford \& Eichler 1987). Interestingly, radio
spectral indices of powerful hotspots are indeed close to --- though not always and
not exactly --- the corresponding value $\alpha_{\rm inj} = (p-1)/2 = 0.5$. Such an agreement
supported the `non-relativistic shock' paradigm for the origin of non-thermal electrons
within heads and cocoons of FR~II type radio sources. It stimulated further development
of this idea in the context of modeling the non-thermal hotspots' emission (Heavens \&
Meisenheimer 1987), enabling the successful explanation of several observational features
(e.g., Meisenheimer et al. 1989, 1997). An important implication is
that in a framework of such a `standard' model the injection spectral index for \emph{all}
the cocoons of powerful radio sources is expected to be strictly $\alpha_{\rm inj} = 0.5$.

As mentioned in section 2.3, a number of low-frequency radio observations of
FR~II sources indicate that there is no universal injection spectral index for
their cocoons (see in this context Rudnick et al. 1994). This conclusion is
supported also by the detailed spectral ageing studies of the few brightest
objects of this type, such as Cyg\,A (Carilli et al. 1991). In addition,
the standard assumption regarding non-relativistic velocities of the
large-scale jets was recently put into question by the most recent X-ray
observations (e.g., Celotti et al. 2001, Tavecchio et al. 2005). If this is indeed the
case, then one should expect significant modifications in the energy spectrum of the
accelerated particles when compared to the non-relativistic limit. Moreover, as argued
below, also the assumption regarding a single power-law form of the injected electrons
is most likely far from being realistic. For these reasons, in our analysis we decided
to treat $\alpha_{\rm inj}$ as a basically free `effective' parameter, which may be
different for different sources, and which may be far from the canonical value of $0.5$.

\subsection{Low-energy electrons}

In this paper we assumed that the power-law continuum of electrons injected by the hotspot
to the cocoons extends down to electron Lorentz factors $\gamma_{\rm i, min} =1$.
In general, synchrotron emission of the electrons with $\gamma_{\rm i} < 10^3$
cannot be observed directly from the hotspots, since it corresponds to frequencies
$< 100$ MHz for a typical value of the hotspots magnetic field $\sim 100$ $\mu$G.
Thus, the assumption regarding the low-energy electron spectrum cannot be verified
by observations in a simple way, and should rather be constrained by the theory of
particle acceleration.

Unfortunately, such a theory is not yet well developed. We note in this context, that
powerful jets at large scales are most likely dynamically dominated by cold
protons (see, e.g., Sikora et al. 2005 and references therein). This means that the
thickness of the terminal reverse shock should be roughly a few gyroradii of the non-
or mildly-relativistic protons, and hence that the energies of the electrons which
are able to undergo diffusive shock acceleration must already be ultrarelativistic,
$\gamma_{\rm i} > m_{\rm p} / m_{\rm e}$. The questions are then (i) which process
is responsible for acceleration of $1 < \gamma_{\rm i} < 10^3 - 10^4$ electrons, (ii)
is the resulting low-energy part of the electron continuum characterized
by a power-law form at all, and if so (iii) is the low-energy spectral index the same
as for the high-energy part of the electron continuum ($\gamma_{\rm i}
> 10^3-10^4$), originating (by assumption) from the diffusive shock acceleration process.

Such problems in the context of FR~II hotspots were not discussed much in the literature,
with only a few papers addressing
these important issues. For example, Kino \& Takahara (2004) analyzed limiting possibilities
for the efficient and non-efficient \emph{temperature coupling} between dynamically dominating
protons and electrons at the jet terminal shock, resulting in the formation of an
ultrarelativistic maxwellian-type electron distribution at energies $\gamma_{\rm i}
\sim \Gamma_{\rm jet} \, m_{\rm p} / m_{\rm e}$ and $\gamma_{\rm i} \sim \Gamma_{\rm jet}$,
respectively (where $\Gamma_{\rm jet}$ is the jet bulk Lorentz factor), followed
by the shock-formed high-energy power-law tail. They argued, that multiwavelength
observations of the Cyg\,A hotspots contradict both possibilities. This implies either an
electron-positron jet composition (as preferred by Kino \& Takahara 2004), or simply more
complex (broad-band) energy spectrum of low-energy electrons (instead of an almost monoenergetical
maxwellian-like distribution). In fact, formation of such a broad-band energy spectrum seems to
be more likely, since in the case of extragalactic jets any energy/temperature coupling between
protons and electrons cannot be collisional in nature, but must be mediated by the magnetic field
and plasma turbulence (see in this context Hoshino et al. 1992).

To sum up, if the dynamical domination of the cold protons within powerful extragalactic jets
is the case, one should expect (possibly significant) differences in the spectral shape
of the low-energy ($1 < \gamma_{\rm i} < 10^3-10^4$) and high-energy ($\gamma_{\rm i} > 10^4$)
parts of the electron continuum formed at jet terminal shocks. Indeed, low-frequency
flattening observed in the radio emission of Cyg\,A hotspots (Carilli et al. 1991), as well
as the most recent analysis of the X-ray emission of the lobes of giant radio galaxy 4C\,39.24
(Blundell et al. 2006), are both consistent with such an expectation. This in turn implies that
the broad-band electron spectrum injected by the hotspots to the expanding cocoons is in reality
not of a simple single power-law form.

\subsection{Relativistic shock velocities}

In the section above we argued that the low-energy part of the electron distribution
formed at the hotspots of powerful jets is most probably broad-band, possibly even
power-law, but with the (unconstrained at the moment) power-law slope most likely
different from the one characterizing its high-energy part. The difference may be due
to the fact that while the low-energy part is formed by poorly known processes of
energy exchange between different plasma species within the colissionless shock front,
the high-energy part is supposed to originate from the `standard' Fermi I
acceleration mechanism. Yet another issue is if this high-energy part always possesses
some universal value of its power-law slope, the same for every powerful radio galaxy. 
We believe not, since even if the diffusive shock acceleration process is
indeed responsible for formation of the latter, the expected relativistic effects
will modify the resulting electron spectrum when compared to the non-relativistic
test-particle model value $p=2$.

As already mentioned, there is growing evidence that relativistic velocities are
present in large-scale FR~II and quasar jets, and so the reverse shocks formed
at the jet heads are at least mildly relativistic. An important fact in this context
is that polarization studies usually indicate magnetic fields perpendicular to the jet
axis within the hotspots (even in the case of giant-sized radio galaxies, see Machalski
et al. 2006), and hence an oblique orientation of the mean magnetic field lines with
respect to the reverse shock front. It was noted by Begelman \& Kirk (1990), that oblique
relativistic shocks are likely to be superluminal, in a sense that the shock velocity
projected on the magnetic field vector exceeds speed of light. In such a
situation, diffusive (Fermi I) acceleration process may not be possible at all,
since particles moving on average along the magnetic field lines are not able
to cross the shock front more than once. Moreover, as shown recently by Niemiec
\& Ostrowski (2004), even if the magnetic field configuration is luckily such
that the diffusive particle scattering at both sides of the (mildly-)relativistic shock
is in principle possible, the resulting energy spectrum of the accelerated particles
strongly depends on the turbulence conditions in vicinity of the shock. As a result,
\emph{a variety} of spectral shapes for the accelerated particles is obtained,
including different spectral indices and deviations from power-law forms.
We believe that this may be partly the reason for different values of the
injection spectral index obtained here for different sources.

\subsection{Jet/hotspot intermittency}

Multiple hotspots detected at the same sides of many FR~II radio galaxies and quasars
indicates that the jet terminal features are \emph{intermittent} in nature (see, e.g.,
Scheuer 1982, Williams \& Gull 1985, Leahy \& Perley 1995). Typically, a pair of the hotspots
is observed at one side of the cocoon, with one feature being stronger and more extended
(`secondary hotspot'), and the other one being weaker and more compact (`primary hotspot'),
as in the famous source Cyg\,A. The secondary and primary hotspots are identified with
`older (extinguishing)' and `younger (recent)' jet termination points. We note, that such
an intermittency may be a direct consequence of the intermittent nature of the jet phenomenon,
discussed in the literature for a number of reasons (e.g., Reynolds \& Begelman 1997,
Stawarz et al. 2004).

Whatever the case is, the issue of interest here is that multiple hotspots differ in
multifrequency radiative and geometrical properties (see Hardcastle et al. 1998).
This should in fact be expected as a natural consequence of the ageing/evolutionary
processes taking place in a mildly-relativistic shock front `fed' by intermittent jet
outflow. An important consequence of this fact is that the electron energy distribution
injected to the cocoons by the hotspots, characterized by lifetimes much shorter
than the lifetime of the whole radio source, is in fact an average over different
activity phases of the jet terminal features. As a result, the injection spectral index
inferred from the analysis of the cocoon's radio continuum may differ substantially
from the low-frequency spectral index of the currently active hotspot in a given
radio source, as suggested in our modeling.\\
\\
The above discussion, pointing out all the reasons why one should not expect a single
broad-band power-law shape of the electrons accelerated at the jet head and injected
in to the expanding cocoons, agrees with the results of detailed investigation
of radio continuum in FR~II sources by, e.g., Rudnick et al. (1994), Treichel et al.
(2001), and Young et al. (2005). These authors, by applying sophisticated technique
of radio data analysis, found that there is no universal injection power-law spectral index
for the cocoons of powerful radio sources, that the injected population of relativistic
electrons thereby is characterized by complex (`curved') spectral shape, and that
there exists a variety of distinct spectral features within the extended lobes. We argue,
however, that for the purpose of simple estimates regarding the sources' main parameters
(such as ages, jet powers, etc.) as presented here, a single power-law injection spectrum
may be a relatively good approximation. The \emph{effective} spectral index of such
a model particle energy distribution may however be different for different sources.
In this context we note that somewhat surprisingly our analysis suggests a relatively
narrow range for this parameter, namely $p = 2.0 - 2.4$.

\section{Conclusions}

Here we propose a new approach to determining the ages of powerful radio sources, on the one hand
exploiting a dynamical model developed for these objects by Kaiser et al. (1997; KDA), and on
the other hand using multifrequency radio observations not necessarily restricted to the
high-resolution ones. The KDA model is chosen for its minimal number of free parameters
when compared to the more sophisticated scenarios by Blundell et al. (1999) or Manolakou
\& Kirk (2002).

We apply the assumed dynamical model to a number of FR~II type radio
galaxies observed at different radio frequencies, and fit --- for each frequency separately
--- the model free parameters to the sources' observed quantities. Such a procedure, enlarging
in fact a number of observables, enables us to determine, relatively precisely, ages and other
crucial characteristics (such as the jet kinetic power) of the analyzed sources. We show,
that the resulting age estimates extend from about 1 Myr for the small-size but powerful
high-redshift radio galaxies to over 100 Myr for the giant-sized radio galaxies. The estimated
ages agree very well with the ones obtained by means of `classical' 
spectral ageing method for objects not older than 10 Myr, for which good-quality data
are available. Meanwhile, in the case of sources older than 10 Myr, our age estimates are
systematically larger (by a factor of $2-4$) than the ages obtained by means of spectral ageing
analysis. Such a behavior is in fact consistent with the conclusions of Blundell \& Rawlings
(2000), who argued that the spectral ageing analysis is most likely erroneous for the lobes
older than 10 Myr.

Interestingly, the ages of FRII-type sources estimated by using the presented method correspond
always to `realistic' values of the jet advance velocity $\leq 0.3\,c$, i.e. consistent with those
measured directly with the VLBI technique in many of the CSO-type sources (see section 5.2.). Thus,
we believe that
the method presented here is applicable also in the case of the oldest radio galaxies, and/or
ones for which the only available low-resolution radio data do not allow for detailed
spectral ageing studies (i.e., for majority of powerful radio sources). This implies
that the applied procedure of favoring the value of the injection spectral index which
corresponds to the minimum of the cocoons' total energy (see section 4.3),
although to some level arbitrary, nevertheless gives very reasonable results.

Our analysis indicates that the main factor precluding precise age determination
for FR~II type radio galaxies regards the poorly known shape of the initial electron energy
distribution injected by the jet terminal shocks to the expanding lobes/cocoons.
We consider briefly this issue, and conclude that the broad-band single power-law form
assumed here (and in all the other analogous models) may be accurate enough for the
presented estimates, although most likely it does not strictly correspond to some well-defined
realistic particle acceleration process. Instead, it should be considered as a simplest model
approximation of the initial electron continuum, averaged over a very broad energy
range and over the age of the source. The \emph{effective} spectral index of such
a distribution may be different for different sources, however within the relatively
narrow range $p = 2.0-2.4$ suggested by our modeling.

In a subsequent paper we indend to apply the presented method for the age determination
in a large sample of FR~II-type (but also younger, MSO-type) radio sources.

\begin{acknowledgements}
This project was supported by MEiN with funding for scientific research in years
2005-2007 under contract No. 0425/P03/2005/29.
\L .S.\ acknowledges also support by the ENIGMA Network through the grant HPRN-CT-2002-00321,
and by MEiN through the grant 1-P03D-003-29 in years 2005-2008.

\end{acknowledgements}

\end{document}